\documentclass[a4paper,10pt]{article}
\usepackage[utf8]{inputenc}
\usepackage{authblk}
\usepackage{graphicx}
\usepackage{amsmath}
\usepackage{amsfonts}
\usepackage{mathtools}
\usepackage{hyperref}
\usepackage{amssymb}
\usepackage{caption}
\usepackage{subcaption}
\usepackage{float}
\usepackage{soul}

\DeclareFontFamily{OT1}{pzc}{}
\DeclareFontShape{OT1}{pzc}{m}{it}{<-> s * [1.0] pzcmi7t}{}
\DeclareMathAlphabet{\mathpzc}{OT1}{pzc}{m}{it}

\usepackage{todonotes}

\DeclareMathOperator*{\argmax}{arg\,max}

\title{ Knowledge transfer across cell lines using Hybrid Gaussian Process models with entity embedding vectors}
\author[1]{Clemens Hutter}
\author[1]{Moritz von Stosch}
\author[1]{Mariano Nicolas Cruz Bournazou}
\author[1]{Alessandro Butt\'e \thanks{Corresponding author. E-mail: a.butte@datahow.ch}}
\affil[1]{DataHow AG, Zurich, Switzerland}

\date{\today}

\begin{document}
\DeclarePairedDelimiter{\norm}{\lVert}{\rVert}
\newcommand*{\R}{\mathbb{R}}

\newcommand{\keywords}[1]{\textbf{Keywords: #1}}

\newcommand{\novelProduct}[0]{\textbf{NP}}
\newcommand{\histProduct}[1]{\textbf{HP#1}}
\newcommand{\nNovelRuns}[0]{N_{\novelProduct}}
\newcommand{\nHistRuns}[0]{N_{\histProduct{}}}
\newcommand{\nProds}[0]{{N_{prods}}}
\newcommand{\nExp}[0]{N_{\Exp{}}}
\newcommand{\nTestExp}[0]{\nExp'}
\newcommand{\nTime}[0]{N_{time}}

\newcommand{\extraFeatFunc}[1]{ExtraFeatures(#1)}

\newcommand{\measMat}[0]{\mathcal{M}}

\newcommand{\nQuant}[0]{{N_{Q}}}

\newcommand*{\regModel}{\Phi}
\newcommand{\ExpSym}[0]{\mathcal{E}}
\newcommand{\Exp}[1]{\ExpSym^{#1}}

\newcommand{\normFeat}[0]{\mathpzc{f}}
\newcommand{\oneHotSym}[0]{e}
\newcommand{\oneHot}[1]{\oneHotSym^{#1}}

\newcommand{\prodID}[0]{\mathpzc{p}}

\newcommand{\prodLS}[0]{\gamma}

\newcommand{\embedDim}[0]{D}
\newcommand{\embedMat}[0]{\mathbf{E}}
\newcommand{\embedVecSym}[0]{E}
\newcommand{\embedVec}[1]{\embedVecSym^{#1}}

\maketitle
 
\begin{abstract}
    {
To date, a large number of experiments are performed to develop a biochemical process. The generated data is used only once, to take decisions for development. Could we exploit data of already developed processes to make predictions for a novel process, we could significantly reduce the number of experiments needed. Processes for different products exhibit differences in behaviour, typically only a subset behave similar. Therefore, effective learning on multiple product spanning process data requires a sensible representation of the product identity.
We propose to represent the product identity (a categorical feature) by  embedding vectors that serve as input to a Gaussian Process regression model. 
We demonstrate how the embedding vectors can be learned from process data and show that they capture an interpretable notion of product similarity. The improvement in performance is compared to traditional one-hot encoding on a simulated cross product learning task. All in all, the proposed method could render possible significant reductions in wet-lab experiments.
    }
    

\end{abstract}


\section{Introduction}\label{sec:intro}
The production of therapeutic proteins and vaccine at scale requires manufacturing processes with a very precise control of the final drug quality, that are economically viable and are outstanding in the protection of environment, health and safety \cite{Lee2015,Yu2017}. The development of these processes is extremely time consuming, highly unreliable and costly, since the process design and conditions need to be tailored to the specific characteristics of the product sought to be produced \cite{Portela2020}. Even for “platform processes”, a large number of process design parameters need to be optimized to increase yields and meet product quality specifications \cite{Li2010}.
Recent years have seen a rise of miniaturized high-throughput platforms (fueled by advances in liquid handling stations) that are representative of some part of the production process, enabling parallel cost- and time-efficient experimentation \cite{Rameez2014,Yellela2018, Shrirao2018, Bushnell2020}.

With the rise of high-throughput technologies \cite{Hans2020}, the process development bottleneck has shifted to i) gathering and analysing the generated experimental data; as well as ii) designing the many parallel experimental runs, such that the data of each run is informative. While statistical Design of Experiment methods have come some way towards addressing this challenge \cite{Politis2017}, these methods largely disregard prior knowledge  about the process at hand \cite{Stosch2014} (in particular when it comes to changes in the cell line), because i) this knowledge is not available in a format that could be explored for experiment design, and ii) though data are produced for similar processes, set-ups for iterative and successive learning (knowledge gathering) from the data are scarce. Hence, horizontal knowledge transfer from product to product, cell line  to cell line, remains limited and entire process design spaces need to be studied \emph{de novo} for every product. This implies that for every new product a similar number of experiments will be required to understand the process behaviour. In previous articles, the possibility of creating more powerful process models has been explored \cite{Narayanan2019,Narayanan2020}, which can be combined with the use of model-based Design of Experiment \cite{Stosch2018,Teixeira2006,Ferreira2014,Brendel2008,Kuchenmueller2020,Abt2018,Anane2019,Hans2020} to reduce the experimental effort. However, for a successive acceleration of process development activities novel methods are needed that allow to derive information from a joint analysis of process data generated for different product and further are even capable to predict process behaviour for novel products, for which yet a limited amount of data has been generated.

To date three approaches for the analysis of multiple product spanning process data are available, as discussed below: one hot encoding (dummy or categorical variables), product specific scaling, or integrating ``handcrafted'' features that describe the characteristics of the product (e.g. molecular descriptors).
\begin{itemize}
\item \emph{One hot encoding}: The product identity for each data point is handled as a categorical process variable and for analysis purposes transformed into a one hot (dummy) variable \cite{Gori2018}. While for analysis purposes it might be efficient to differentiate between data of different products, the learnings that can be transferred from one product to the next remain limited, since the one hot encoding does not carry information about the similarity of the different products. Furthermore, it is not possible to filter and extract only the information that is useful for the new product.

\item \emph{Scaling}: Scaling can improve the comparability of process data from different products \cite{Li2017,Liu2019}. One can use, for instance, in process scaling to ``normalise'' for differences in absolute titers or to understand the impact of pH-variations around different set-points. However, if the scaling does not result in making the data more comparable then a joint analysis will likely provide as much insight as an analysis of the data of each product alone.

\item \emph{Handcrafted, knowledge-based feature creation}: Integrating features into the analysis that describe the product characteristics can enable cross-product data analysis. To this end, molecular descriptors have been generated and used, giving rise to quantitative structure–activity relationship (QSAR) models, quantitative sequence-activity modelling (QSAM) or convolution graph networks \cite{Schweidtmann2020}. Karlberg et al \cite{Karlberg2018} proposed to use mAb structure characteristics for a streamlined implementation of the Quality by Design (QbD) paradigm. However, it is not straightforward to capture the characteristics of a drug substance and additionally select the features of interest from the multitude of generated features for a limited amount of data is also not trivial.
\end{itemize}

In this work we demonstrate for the first time the potential of product embeddings with Gaussian Process Regression for multiple product spanning process data modeling in bioprocess development. 
The concept of embedding categorical entities (products in our case) was inspired by the representation of words with high dimensional vectors in the field of natural language processing. These embeddings have many useful characteristics like locating words with similar meaning close by \cite{glove}. They can even exhibit \emph{additive compositionality} such that the embedding of the word ``Berlin'' is closest to the sum of the embedding vectors for ``capital'' and ``Germany'' \cite{word2vec}. Usually, word embeddings are learned as weights of a neural network on a generic task \cite{Mikolov2013}. 
In bioprocess modeling, Gaussian Process regression has notable advantages such as lower data requirements and the possibility to assess the uncertainty of predictions. We hence propose a method to learn embedding vectors with this regression algorithm from bioprocess data.

Using the embeddings, we seek to explain the similarities in the biological system across products. Variations sourcing from process operations are explicitly described using material balances, see e.g. \cite{Richelle2020} for more details. The resulting system of equations naturally constitutes a hybrid semi-parametric model. 

In what follows, we first describe the proposed approach to learn product embeddings with Gaussian Process models integrated into a hybrid semi-parametric process model. We derive mathematical insight into the shortcomings of the traditional product representation with one-hot vectors in Gaussian Process regression and show how our learned embeddings provide insight into similarities between products. We then rigorously evaluate the proposed approach using a simulation case study and finalize by summarizing the findings.

\section{Material and Methods}\label{sec:matrials}

\subsection{Bioprocess Data in Process development}\label{sec:data_structure}

In the development of a process the aim is to find process conditions (e.g. temperatures, initial concentrations, etc.) that consistently produce a large amount of high quality titer (product concentration) in a short amount of time. 
To effectively support the development of a process for a novel product, we want a model that can predict the system behaviour for an unseen process condition $\ExpSym'$, such that we can use the model for process optimization.
The behavior of the system is here represented by a matrix of concentrations $\measMat' \in \R^{\nTime \times \nQuant}$ for $\nQuant$ relevant quantities at $\nTime$ time points throughout the process evolution. 

To create such a model, data from $\nExp$ experimental runs which had been generated in the past for $\nProds$ different products could be used, as we show in the following. For each run $i \in \{1 \dots \nExp\}$, the process condition $\Exp{i}$ and the product $\prodID^i \in \{ 1, \dots, \nProds \}$ as well as the matrix of concentrations $\measMat^i \in \R^{\nTime \times \nQuant}$ are considered to be known. 
\subsection{Hybrid Regression Models}\label{sec:stepwise}

\newcommand{\nExtraFeat}[0]{k}
The modeling of the dynamic evolution of the concentrations is inspired by the dynamic material balances of an ideally mixed reactor.  Specifically, by discretizing in time, we approximate for each experiment in the training set $i \in \{1, \dots, \nExp\}$ the slope of the change in concentrations between each consecutive pair of measurements as follows
\begin{equation}
y_t^i := \frac{\measMat_{t+1}^i - \measMat_{t}^i}{\Delta t} 
\end{equation}
where $\measMat^i_t \in \R^\nQuant$ is the vector of measurements at time $t$ and $\Delta t$ is the time between two measurements. Further, we construct feature vectors $x_t^i$ by concatenating the concentrations at the beginning of the step $\measMat^i_t$ with additional features $\extraFeatFunc{\Exp{i}} \in \R^\nExtraFeat{}$ such as reactor temperature and pH which are obtained from the experimental design. Doing this for all experiments in the training set we obtain matrices $X\in \R^{(\nExp \cdot \nTime) \times (\nQuant + \nExtraFeat{})}$ and $Y \in \R^{(\nExp \cdot \nTime) \times \nQuant}$. We use these to train an arbitrary regression model $\regModel: \R^{\nQuant + \nExtraFeat{}} \rightarrow \R^\nQuant$, which describes the reaction kinetics in the reactor analogy.

The prediction for an unseen experimental condition $\ExpSym{}'$ is given by 
\begin{align}
    \measMat'_0 &= InitialConcentration(\ExpSym{}' ) \\
    \measMat'_{t+1} &= \measMat'_t + \regModel\left((\measMat'_t \quad ExtraFeatures(\ExpSym{}') ) \right)\cdot \Delta t
\end{align}

Additionally, the effects of mass feeds (such as Glucose and Glutamine in the presented simulation case) can be accounted for explicitly using the mechanistic mass balance calculations, see e.g. \cite{Richelle2020}.

\subsection{Gaussian Process Regression}\label{sec:gp}
Gaussian Process regression (GP) is a particular choice for the class of regression model used in section \ref{sec:stepwise}. It approximates a function $f: \R^d \rightarrow \R$ where $d$ is the number of features. In this brief outline we assume for simplicity that there is no measurement noise and refer to \cite{RW} for more details. A GP regression model requires a kernel or covariance function $k(\cdot, \cdot): \R^d \times \R^d \rightarrow \R$ which should capture a notion of similarity that is appropriate for the application \cite[Chapter 4]{RW}. Using the kernel, a prior over the function $f(\cdot)$ is defined. The prior specifies that the function values $\overline{Y}\in \R^{n+1}$ evaluated at a set of $n+1$ points $\overline{X}\in\R^{(n+1) \times d}$ follow a normal distribution 

\begin{align}\label{eq:gp_prior}
\overline{Y} \thicksim \mathcal{N}(0, K(\overline{X}, \overline{X}))
\end{align} 

where the matrix $K(\overline{X}, \overline{X})\in \R^{(n+1) \times (n+1)}$ is given by applying the kernel to each pair of rows: $K(\overline{X}, \overline{X})_{ij} = k(\overline{x}_i, \overline{x}_j)$. We can treat $n$ of these datapoints as observed (i.e. training data $X, Y$) and the remaining point as a query point for which only $x^*$ is known. A predictive distribution over $f(x^*)=y^*$ is then obtained by conditioning the prior with the observed training data \cite[Section 2.2]{RW}: 

\begin{align}
\begin{split}\label{eq:gp_pred}
    y^* \mid x^*, X, Y \thicksim \mathcal{N}(& K( x^*, X)K(X, X)^{-1}Y,\\
    &k(x^*, x^*)- K(x^*, X)K(X, X)^{-1}K( X, x^*))
\end{split}
\end{align}

Among the most widely used kernels is the squared exponential kernel (or radial basis function) with automatic relevance determination (ARD)
\begin{equation}\label{eq:rbf}
    RBF_{\theta}(x, x') = exp \left(-\sum_{i=1}^d \frac{(x_i - x'_i)^2}{2\theta_i ^2} \right)
\end{equation}
where $\theta \in \R^d$ is a vector of hyper-parameters. For the RBF kernel these parameters are usually called length-scales. When using a parameterised kernel, the prior (\ref{eq:gp_prior}) also depends on $\theta$. 
The value of $\theta$ can be chosen by maximising the likelihood of the observed training data under the prior:
\begin{equation}\label{eq:find_theta}
\theta :=  \argmax_{\theta'} log P(Y | X, \theta')
\end{equation}
See \cite[section 5.4.1]{RW} for more details. 

The one dimensional regression model can easily be extended to functions $\R^d \rightarrow \R^t$. Prediction is then effected (for fixed $\theta$) by using equation (\ref{eq:gp_pred}) $t$ times independently on each of the $t$ targets. To find $\theta$ one can maximise $\sum_{i=1}^t log P(Y^i \mid X, \theta) $, where $Y^i\in \R^n$ is the vector of the $i$-th target for each of the $n$ training samples. 

\subsection{Representing the different products}
In bioprocess applications, we want to train a GP on data originating from processes of \emph{different} products, which behave differently. Therefore, we need to add a categorical feature to the GP input, that represents from which of the $\nProds$ products the data point was generated. 

\subsubsection{Traditional one-hot representation}\label{sec:one_hot_intro}
Traditionally, categorical features are represented by a one-hot vector (similar to dummy variables in statistics) $\oneHot{\prodID} \in \{0, 1\}^\nProds$ for $\prodID \in \{1, \dots, \nProds\}$ which has a single one at the position corresponding to the product index $\prodID$. For example the third product would be represented by the vector $\oneHot{3} = (0 \quad 0 \quad 1 \quad 0 \quad 0 \quad 0)$. The one-hot vector is then appended to the other features. Hence the input feature vector $x= (\normFeat \quad \oneHotSym)$ to the GP in section \ref{sec:stepwise} is comprised of process state features $\normFeat \in \R^{d'}$ (i.e. current concentration measurements, pH, temperature ...) and the one hot encoding $\oneHotSym$ of the product. 

For a better grasp of the implications we split the sum in (\ref{eq:rbf}) to treat the features and the product representation separately:

\begin{align}
\begin{split}\label{eq:rbf_seperates}
    RBF_\theta(x, x') &= exp \left(-\sum_{i=1}^{d'} \frac{(\normFeat_i - \normFeat'_i)^2}{2\theta_i ^2} - \sum_{j=1}^{\nProds} \frac{(\oneHotSym_j - \oneHotSym'_j)^2}{2\theta_{(d'+j)} ^2} \right) \\
    &=RBF_\theta(\normFeat, \normFeat') \cdot exp \left (- \sum_{j=1}^{\nProds}  \frac{(\oneHotSym_j - \oneHotSym'_j)^2}{2\theta_{(d'+j)} ^2} \right) \\
    &= RBF_\theta(\normFeat, \normFeat') \cdot RBF_{\theta}(\oneHotSym, \oneHotSym')
\end{split}
\end{align} 
where $\oneHotSym$ and $\oneHotSym'$ are the one-hot representations of the two products that data points $x$ and $x'$ were generated from.

Hence the kernel has a large value \emph{iff} both the kernel on $(\normFeat, \normFeat')$ is large \emph{and} the kernel on $(\oneHotSym, \oneHotSym')$ is large. Or in words: $x$ and $x'$ are considered similar \emph{iff} the current concentrations and process conditions are similar \emph{and} the two points come from similar products as measured by $RBF_{\theta}(\oneHotSym, \oneHotSym')$. 

\subsubsection{Traditional representation cannot capture pairwise similarities} \label{sec:one_hot_bad}
In the following we argue that the traditional one-hot representation with an ARD-RBF kernel cannot capture useful similarities between products. The length scales can encode how \emph{peculiar} a given product is compared to all other products. However, they cannot encode clustered pairwise similarities between products. 

For notional convenience we define $\prodLS_j := \frac{1}{2\theta_{d'+j}^2} > 0$ and let $\prodID \text{ and } \prodID' \in \{1, \dots, \nProds\}$ be the id of the product that datapoint $x$ and $x'$ respectively came from. Due to the one-hot nature of $\oneHotSym$ the second term in (\ref{eq:rbf_seperates}) can be written as
\begin{equation}
\label{eq:prodKernel}
    RBF_{\prodLS}(\oneHotSym, \oneHotSym') = 
    \begin{cases}
    1 & \mbox{if} \quad \prodID = \prodID'  \\
    exp(-\prodLS_{\prodID} - \prodLS_{\prodID'}) & \mbox{if} \quad \prodID \neq \prodID'
    \end{cases}
\end{equation}

Now consider a thought experiment: Assume we have data from $\nProds=100$ different products that form two clusters. Product 1 and 2 in cluster A are very similar to each other but different from products 3 to 100 which form cluster B. Since the products 3 to 100 are very similar to each other, the kernel in (\ref{eq:prodKernel}) should return a large value for any two products from this cluster. This in turn requires that $\prodLS_3, \dots, \prodLS_{100}$ are small. Further, the kernel similarity should be small for any product from cluster A with any product from cluster B, which implies that $\prodLS_1, \prodLS_2$ have to be large. Consequently, the kernel similarity between products 1 and 2 is small. This is not in line with the assumption of the thought experiment. Hence, even for this simple example, the kernel cannot correctly capture the clustered notion of similarity we desire. 

\subsubsection{The Novel approach - Product embedding}\label{sec:product_embed}
We propose a novel approach to alleviate this issue. We borrow the concept of (word) \emph{embedding vectors} from natural language understanding where each word is represented by a vector which is used as input to other machine learning models (e.g. a neural network). \\
Similarly we propose to represent each product $ \prodID \in \{1 \dots \nProds\}$ by a vector $\embedVec{\prodID} \in \R^{\embedDim}$, which are organised for convenience as columns in a matrix $\embedMat \in \R^{ \embedDim \times \nProds}$. Hence each product is represented by a point in $\embedDim$-dimensional space.

In the input to the GP the one-hot vector $\oneHot{\prodID}$ from the previous approach is now replaced with the respective embedding vector $\embedVec{\prodID}$. For example all feature vectors in the training set that come from the third product would have their last $\embedDim$ entries equal to the third column of the matrix $\embedMat$ (i.e. the third embedding vector). 
So how will the RBF kernel behave with this product representation?

Without losing flexibility we can fix the length scales $\theta$ to 1 for the feature-dimensions that contain the product embedding (instead of changing a length scale we simply scale the corresponding row in $\embedMat$). Hence, the similarity kernel on the product identity (second term in (\ref{eq:rbf_seperates})) becomes 
\begin{equation}
    RBF_1(\embedVecSym{}, \embedVecSym') = exp \left(-\sum_{i=1}^{\embedDim} \frac{(\embedVecSym{}_i - \embedVecSym'_i)^2}{2 \cdot 1} \right ) = exp\left(- \frac{1}{2} \norm{\embedVecSym - \embedVecSym'}^2 \right)
\end{equation}
where $\norm{\cdot}$ is the euclidean norm. With this kernel two products are considered similar (high kernel value) \emph{iff} the two associated points in embedding space are close to each other as measured by euclidean distance. Therefore, 
almost any clustered similarity structure of products can be modelled by choosing appropriate embedding points in a sufficiently high dimensional space. For example, the structure described in the thought experiment in section \ref{sec:one_hot_bad} can readily be expressed with $\embedDim=1$. Note, that the embedding vectors are invariant to rotations, shifts and mirroring. That is, modifying all embedding by $\widetilde{\embedVec{i}} := W \embedVec{i} + b$ with $W\in\R^{\embedDim \times \embedDim}$ orthogonal ($W^T W = \mathbb{I}$) and $b\in \R^{\embedDim}$ does not change kernel values or indeed the predictions of the GP. 

\subsubsection{Identifying the product embedding vectors}\label{sec:product_embed_parameter}
Of course it is usually not possible or convenient to define a suitable embedding for the products manually but they can be determined in the optimization of the hyperparameters $\theta$ using the training data. Specifically, the prior in (\ref{eq:gp_prior}) now depends on $\embedMat$. Hence, in analogy to (\ref{eq:find_theta}), we choose $\embedMat$ together with the length scale vector $\theta$ for the process state features according to 
\begin{equation}\label{eq:find_embbed}
    \theta, \embedMat = \argmax_{\theta', \embedMat'} log P(Y | X, \theta', \embedMat')
\end{equation}

A convenient way to implement this is by defining a custom kernel
with optimisable hyperparameters $\theta, \embedMat$ as
\begin{equation}
    k_{\theta, \embedMat}(x, x') = RBF_{\theta}(\normFeat, \normFeat') \cdot exp \left ( -\frac{1}{2} \norm{\embedMat \oneHotSym - \embedMat \oneHotSym'}^2 \right )
\end{equation}
where $\oneHotSym$ and $\oneHotSym'$ are one-hot representations of the products which are append to the training data as in section \ref{sec:one_hot_intro}. Note that $\embedMat \oneHotSym$ simply picks the column corresponding to the 1 in $\oneHotSym$ from the matrix $\embedMat$. This kernel\footnote{
An implementation can be found at \url{https://github.com/rauwuckl/EntityEmbedding4GP.git}
}
can then be plugged into an existing implementation  of the GP hyperparameter tuning and prediction algorithm (e.g. scikit-learn \cite{scikit-learn, sklearn_api}). 

\subsection{Data Generation with an Emulated Bioprocess} \label{sec:insilico}
Data for $\nProds$ different products is generated using the emulator model described in \cite{Harini2020}. Each product has different values for the parameters of the model equations, such that the process of each product has a characteristic behavior, different to the processes of the other products. To make the process behavior more realistic we allow lactate to be consumed in the processes of some of the products, while it can only be produced in others. 
We draw experimental designs from a Latin Hyper Cube Design (\cite{lhs1, lhs2}) and simulate experimental runs with the emulator for these designs on the different products (i.e. the experimental design is identical across the products). Concentration measurements of VCD, Lactate, Glutamine, Glucose, Ammonia and Titer are performed once a day in the simulation (these are used to construct the matrix $\measMat^i$ of measurements in the simulation case) and corrupted with small additive Gaussian noise to mimic the analytic measurement error. 

\subsection{Performance Metric} 
A test set of $\nTestExp$ unseen experimental conditions is set apart to assess the prediction quality of the model. The absolute error between predicted concentrations and actual measurements are computed to obtain a $\nTime \times \nQuant \times \nTestExp$ tensor. To standardise, we divide the absolute errors by the variance in the actual measurements across the $\nTestExp$ experiments for each time point and quantity. We then take the median across experiments, obtaining a $\nTime \times \nQuant$ matrix. Values well below 1.0 indicate good performance whereas values above 1.0 show that a dummy prediction of the time and quantity dependent mean value would outperform the machine learning model. To provide a rapid understanding of the results we take the average across time to obtain $\nQuant$ accuracy values for the measured quantities. 

\section{Results}\label{sec:results} 
Wet-lab experiments are cost and time expensive, consequently we are interested in using a model in combination with as few experiments as possible to develop the process for a new product. In particular, we are interested in improving model accuracy when there is only data from very few experimental runs available. We conjecture that including historic data from other products can increase model accuracy in prediction for the new product. We investigate this by using a case study with simulated data (section \ref{sec:insilico}): We have 5 \emph{historic} products (\histProduct{1} ... \histProduct{5}) for which we generate data from 16 experimental runs each. Additionally, we have a novel product \novelProduct{} for which we have data from only $\nNovelRuns{} \ll 16$ experimental conditions. We always measure performance of the trained model by predicting 100 unseen experimental conditions of the new product $\novelProduct{}$. 

\subsection{Choosing the Embedding Dimension}


\begin{figure}

     \centering
     \begin{subfigure}[b]{0.50\textwidth}
         \centering
         \caption{Objective (\ref{eq:find_embbed}) on example training set}
         \includegraphics[width=\textwidth]{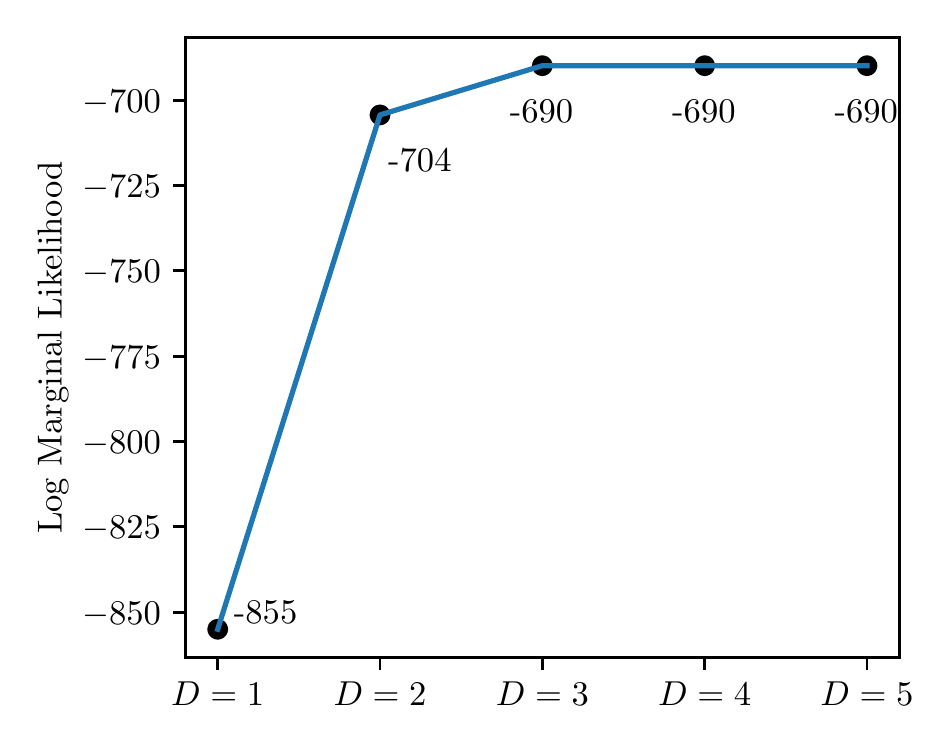}
        \label{fig:choose_dim_practice}
     \end{subfigure}
     \hfill
     \begin{subfigure}[b]{0.44154929577464785\linewidth}
         \centering
         \caption{Test Error Distribution}
         \includegraphics[width=\textwidth]{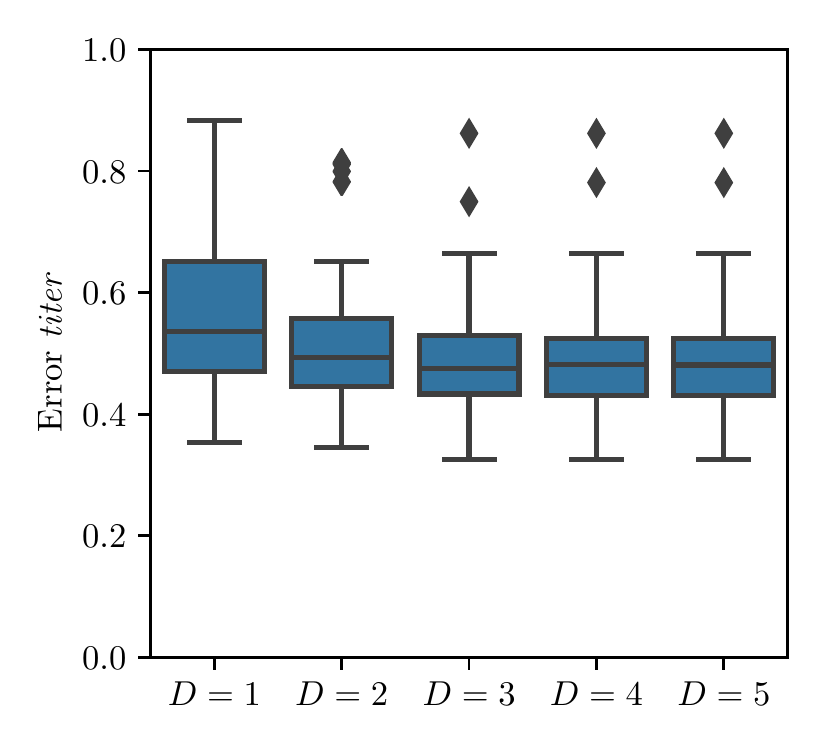}
        \label{fig:choose_dim}       
     \end{subfigure}
        \caption{\textbf{(a)} In practice the highest log marginal likelihood (\ref{eq:find_embbed}) of the training set can be used to select $\embedDim = 3$. \textbf{(b)} We validated this choice by showing the test error distribution obtained with 75 different training and test sets. It confirms that the test error does not improve after $\embedDim = 3$.
        }
        \label{fig:three graphs}
\end{figure}

The embedding dimension $\embedDim$ is chosen based on all data available for training. It has to be large enough such that the appropriate pairwise distances between products can be realised by points in $\embedDim$ dimensional space. Hence we choose the value of $\embedDim$ at which the log marginal likelihood (\ref{eq:find_embbed}) does not significantly increase any more, which indicates that the optimal pairwise distances can already be realised in this space. Figure \ref{fig:choose_dim_practice} shows the likelihood values for an example data set (here $\nNovelRuns = 4$)\footnote{The example dataset is provided at \url{https://github.com/rauwuckl/BioprocessExampleData}.}.
From this we select $\embedDim=3$ because a value of $4$ does not further increase the objective.  

To show that this way of choosing the embedding dimension is valid, we evaluated the choice using a statistical approach. We independently generate 75 pairs of training ($16 \cdot 5 + 4$ runs) and test ($100$ runs) sets. Specifically, the experimental conditions for these sets are independently drawn from a Latin Hypercube Design and the processes are then simulated according to section \ref{sec:insilico}. Hence we obtain 75 independent estimates of the test error. Figure \ref{fig:choose_dim} shows a box plot comparing the distribution of test error for different choices of $\embedDim$. We observe that the error decreases until $\embedDim = 3$ which corroborates that this is actually the choice of $\embedDim$ with the lowest expected test error. 


\subsection{Product Embedding improves Accuracy}

\begin{figure}[t]
\includegraphics[width=\textwidth]{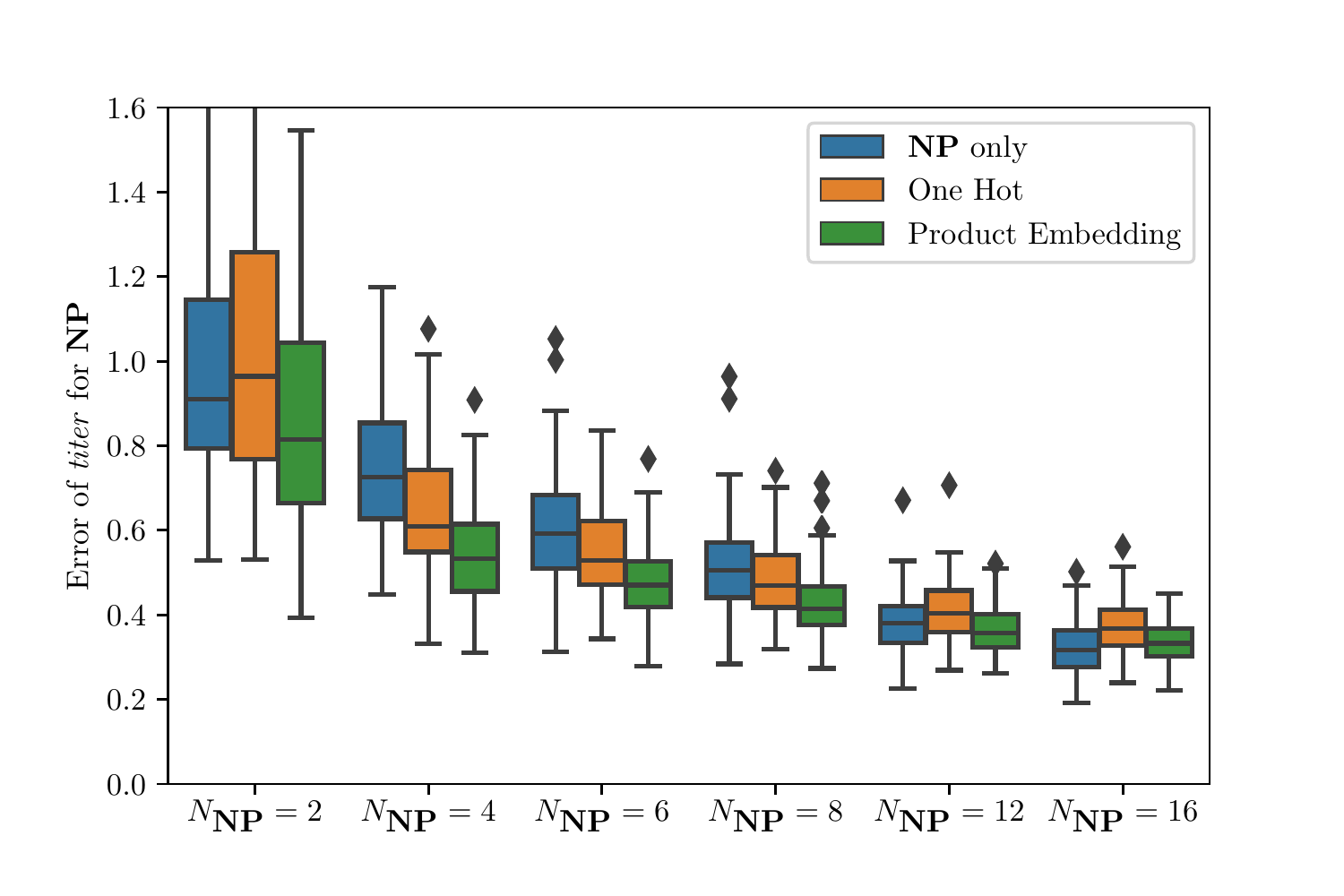}
\caption{The test error distribution for the product \novelProduct{} across 150 random draws of the training set. We vary  $\nNovelRuns$, the number of runs on the product $\novelProduct$ available for training, and try three different algorithms, i.e.: \emph{Product Embedding} is the methodology proposed in section \ref{sec:product_embed}; \emph{One Hot} is the traditional one-hot representation of the product (section \ref{sec:one_hot_intro}) and for \emph{\novelProduct{} only} we only use the data of the $\nNovelRuns$ runs available for the novel product. }
\label{fig:different_k}      
\end{figure}

The performance of the proposed method is evaluated by predicting test experiments of a novel product \novelProduct{}. The models have been trained on data from 16 runs for each of 5 historic products plus data from a varying number $\nNovelRuns$ of runs with the new product. 
We compare the proposed \emph{product embedding} methodology (section \ref{sec:product_embed}) with a traditional one hot representation of products (section \ref{sec:one_hot_intro}). Additionally, we include the baseline \emph{\novelProduct{} only} in which a GP is trained with only the $\nNovelRuns$ available runs of the novel product. The data for the last baseline therefore has less features, because the representation of the product identity is not needed. We fix a test set of 100 runs on the novel product \novelProduct{} and repeat the comparison 150 times with independently generated training sets. This is to ensure statistic validity and avoid reporting results that are merely a consequence of the stochastic variation in one particular training set. Figure \ref{fig:different_k} shows bar plots of the test error distributions. One can clearly see the benefit of adding data from other historic products when there are only few runs ($\nNovelRuns = 4, 6, 8$) of the product of interest available for training, i.e \emph{Product Embedding} and \emph{One Hot} (both using historic data) achieve a lower error then \emph{\novelProduct{} only}. Furthermore, the product embedding approach clearly outperforms the one-hot representation of products. For example, with only $\nNovelRuns = 4$ experimental runs of the novel product the embedding already achieves similar accuracy as the \emph{\novelProduct{} only} baseline with $8$ runs or the one hot representation with $6$ runs.

\begin{figure}[h]
\includegraphics[width=\textwidth]{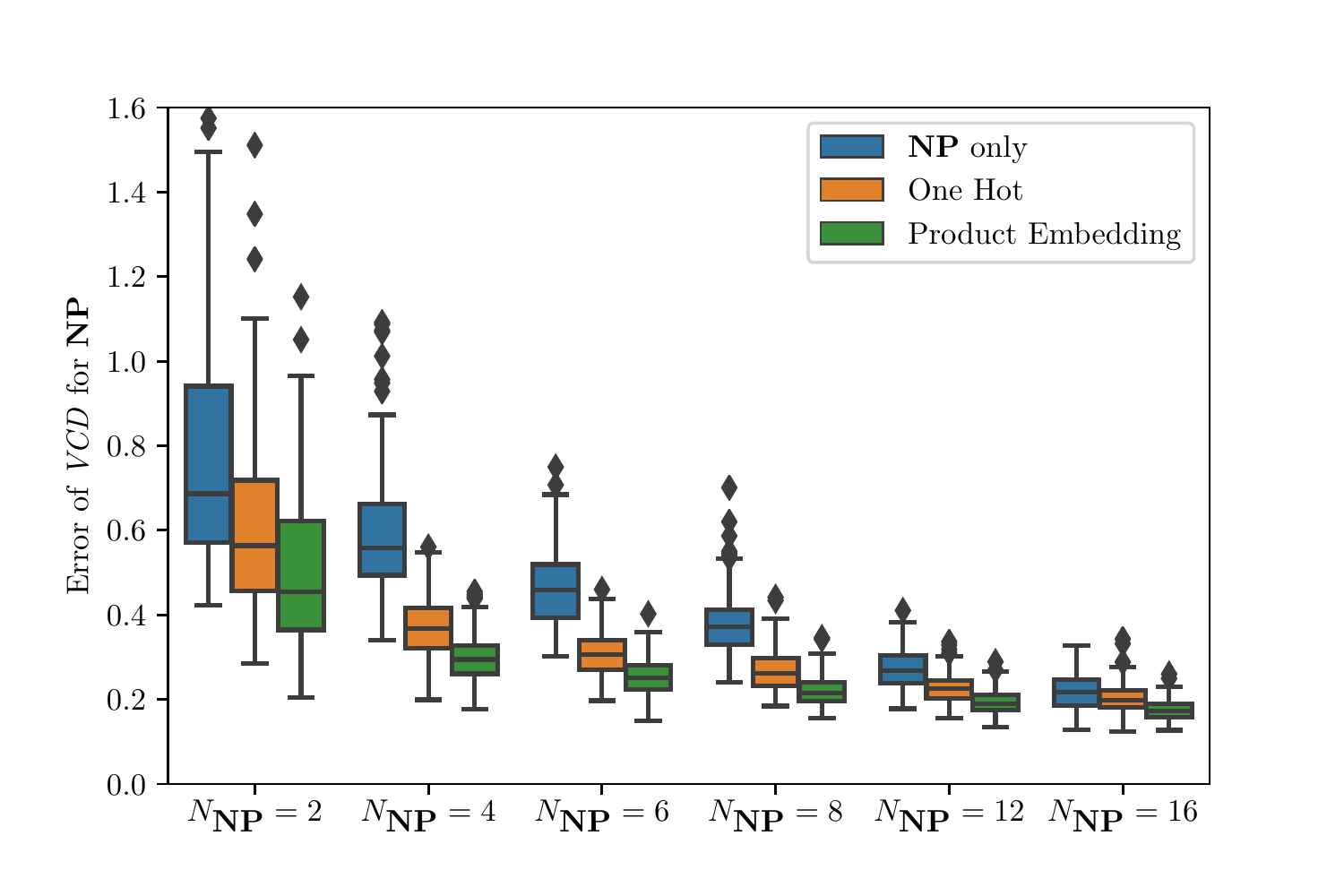}
\caption{Test error distribution over 150 draws of the training set for predicting \emph{Viable Cell Density} concentrations.}
\label{fig:different_k_vcd}      
\end{figure}

As the number of runs of the novel product ($\nNovelRuns{} = 12, 16$) increases, we expect a smaller benefit of using the product embedding with historic data. If enough data for \novelProduct{} is available the product's behaviour can directly be modelled sufficiently well without the necessity to extend the data set with related products.

However, it is surprising that for $\nNovelRuns{}=16$ the \emph{\novelProduct{} only} baseline in fact outperforms the embedding algorithm when predicting \emph{titer}. We speculate that this is because the embedding algorithm focuses on modeling the underlying biological system (i.e. the five interrelated quantities \emph{VCD, Glucose, Glutamine, Lactate, Ammonia}) rather then \emph{titer} production, which has no causal influence on the system dynamics. Hence, the embedding would be biased towards grouping products with similar cell characteristics.
However, two processes of different products, which exhibit similar cell behaviour (and hence close embedding vectors) could (in our emulator model) still produce \emph{titer} at different levels. Here the similarity information from the embedding would actually be slightly detrimental for performance since it would lead to falsely predicting similar \emph{titer} production as well. This could explain the better performance of the \emph{\novelProduct{} only} algorithm specifically for \emph{titer} prediction. 

The performance of \emph{viable cell density} prediction supports this hypothesis (see Figure \ref{fig:different_k_vcd}). For this quantity the \emph{Product Embedding} outperforms the baseline for all values of $\nNovelRuns{}$ which indicates that the selected embeddings are indeed more beneficial for reflecting cell characteristics than titer formation. This could possibly be mitigated by having separate embeddings for the task of predicting \emph{titer}. Nonetheless, for both \emph{titer} and \emph{VCD} we see the largest benefit of the product embedding algorithm when there is few data (i.e. $\nNovelRuns{}=4, 6, 8$).

From the results of the case study the advantage of exploiting historic data with the embedding method for process development becomes evident. It enables us to use an accurate model for making process decisions after just a few experiments. Therefore, the embedding approach can help to reduce the requirement for wet-lab experiments with the novel product, accelerating process development while decreasing costs.

\subsection{Interpretable Embedding Vectors}

\begin{figure}[ht]
  \includegraphics[width=\textwidth]{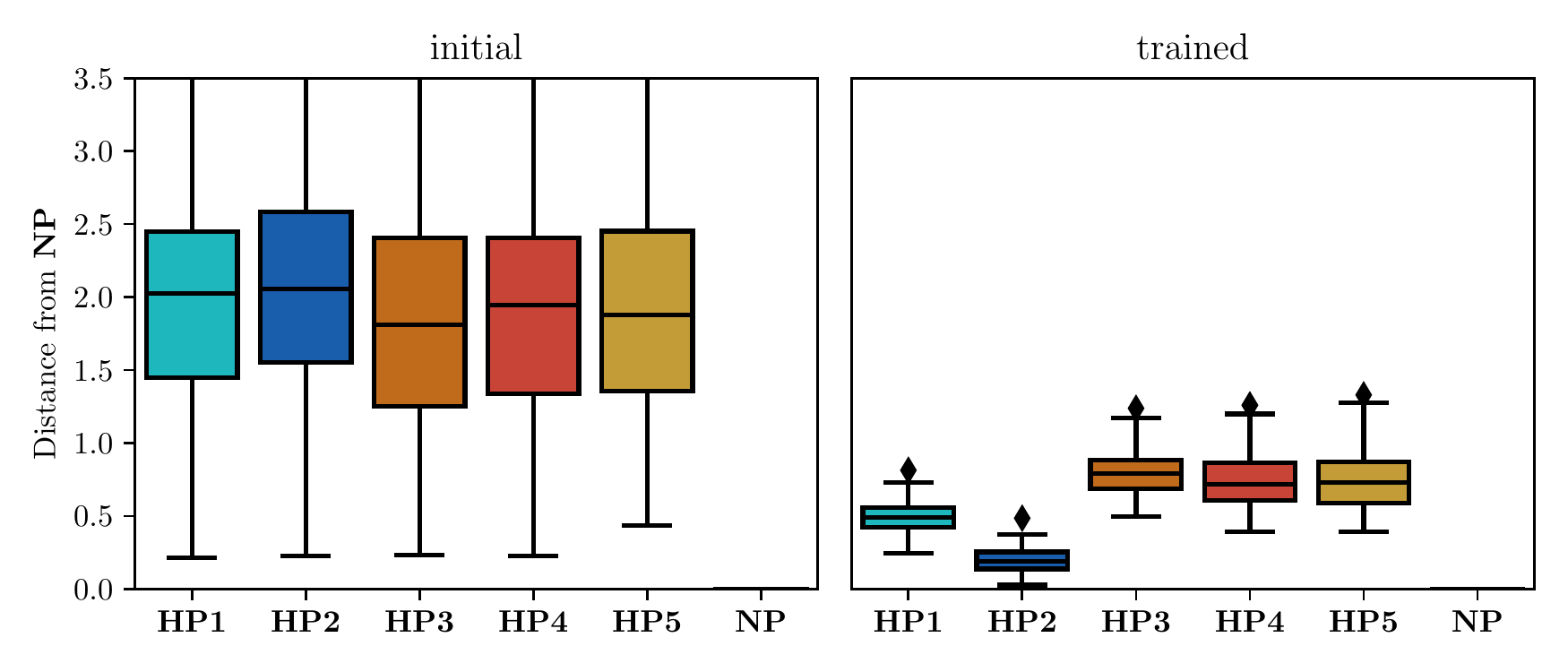}
\caption{Distribution of distances from the product \novelProduct{} to \histProduct{1}\dots{}\histProduct{5} across 150 repetitions.}
\label{fig:dist_curious_pretzel}       
\end{figure}

Each embedding vector is initialised by drawing a point uniformly at random in the cube $[0.1 \quad 3]^3$. During training of the GP the  \emph{L-BFGS-B} \cite{lbfgs}
algorithm is used with this initialisation to find the (possibly local) optimum of the objective (\ref{eq:find_embbed}). 

In section \ref{sec:product_embed} we reasoned that products are similar if the associated embedding points have small euclidean distance. Here, we are particular interested in the distances between the novel product \novelProduct{} and each of the historic products \histProduct{1}, \dots \histProduct{5}. Figure \ref{fig:dist_curious_pretzel} shows the distribution of distances in the embedding space before (left) and after (right) training. Clearly the distances of the trained embeddings are very different from the random initialisation. In particular the small inter quartile ranges indicate that the embedding captures similarity between products in a reliable way that is invariant to the training set used. That is, across the 150 training sets there is only slight variation in the distances between the learned embedding vectors. 

Specifically, we see that products \histProduct{1} and in particular \histProduct{2} are most similar to $\novelProduct$. This is not surprising as these three products all have lactate consuming behaviour whereas products \histProduct{3}, \histProduct{4} and \histProduct{5} do not. 

\begin{figure}[ht]
  \includegraphics[width=\textwidth]{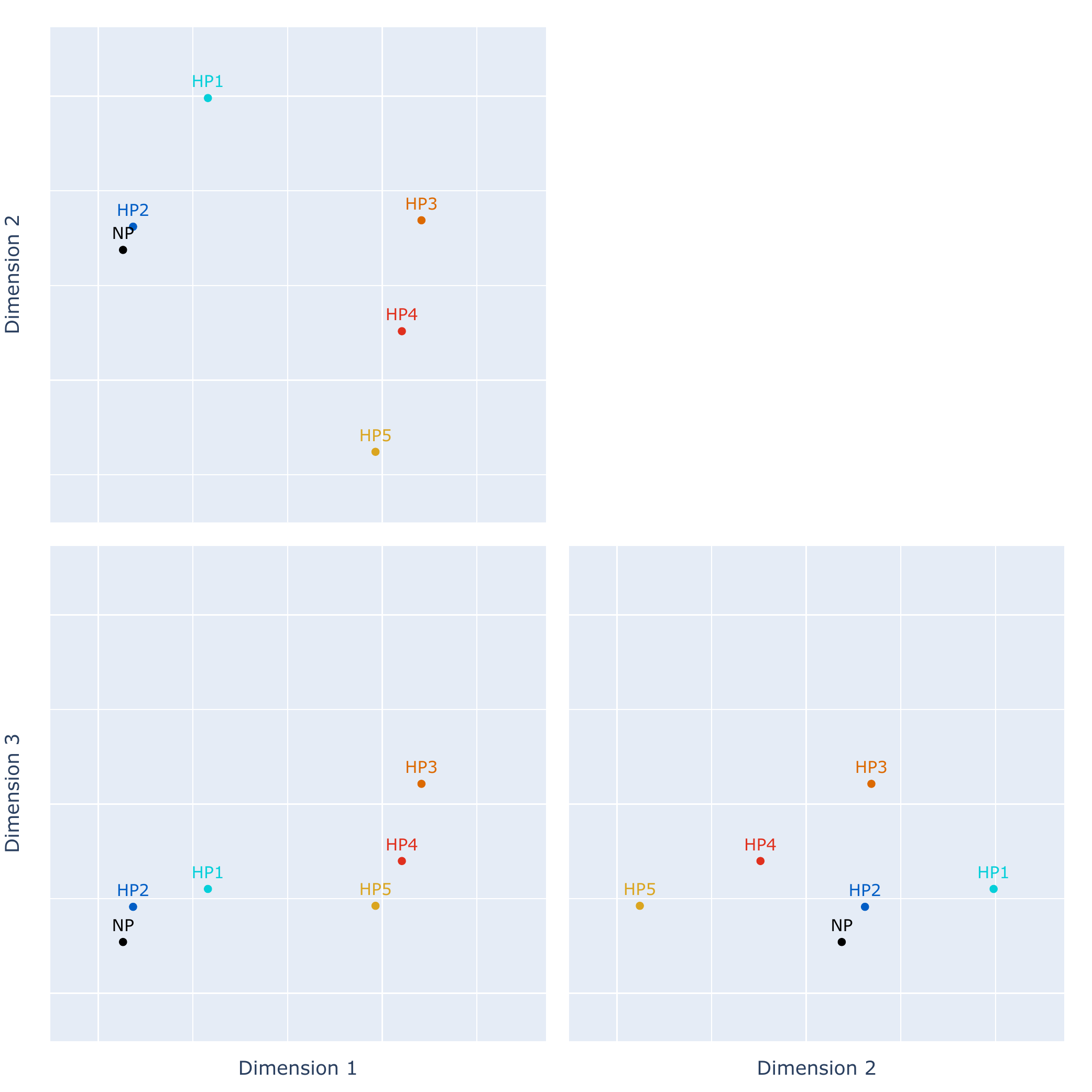}
\caption{One example of a product embedding.  
}
\label{fig:embed_example}       
\end{figure}

The similarity structure observed in Figure \ref{fig:dist_curious_pretzel}, can also be found by looking at the product embedding vectors learned from one data set (Figure \ref{fig:embed_example}). 
This embedding was obtained using the example training set (same as Figure \ref{fig:choose_dim_practice}) and hence represents a practical application of the algorithm with a real data set, i.e. as when using one data set during process development. It demonstrates that the lactate consuming products \histProduct{1}, and $\novelProduct$ form a visually inferred group around \histProduct{2}, whereas the products \histProduct{3} and \histProduct{5} form a looser group around \histProduct{4}. We also clearly see that $\histProduct{2}$ is very similar to the novel $\novelProduct{}$. 
In the development of a new process such insights can help to understand the process behavior of the novel product and its relationship to historic, well studied processes. 
This itself is a valuable use case of the product embedding algorithm in addition to the improved predictive accuracy.

The qualitative structure exemplified by Figure \ref{fig:embed_example} is statistically relevant and found in most of the 150 repetitions. To confirm this, we include in the appendix plots similar to Figure \ref{fig:dist_curious_pretzel} that show the distribution of distances from each of the 5 historic products. 

\section{Conclusions}
We propose an embedding approach for multiple product spanning probabilistic process data modeling and evaluate it using an upstream mammalian (cell-culture) bioprocess simulation case study. In particular, we use a hybrid regression model to predict the evolution of the bioprocesses. 

We demonstrate that prediction performance of the hybrid model can be improved by including data from other, historic products, using the proposed embedding approach. This creates a model capable of transferring knowledge from data of processes which behave similar and thereby reduces the data required with the new process to train an accurate model. 
We show that visualising the embedding space can offer valuable insights for experts about the complex relation between products and processes. More importantly, during prediction it enables the algorithm to give higher importance to information from those data points in the training set that were obtained with similar behaving products. For this reason the embedding algorithm outperforms the traditional one-hot representation, as was shown with the simulation study. 

The findings are highly useful in process development where one aims to find process conditions that produce a desired outcome (e.g. high \emph{titer} production). The model can already be trained to high accuracy after obtaining data from only few experiments with the novel process by leveraging available historical data from different processes. For instance, it has been shown that with only four experiments on the new product, one could reach a similar model precision to the case where eight to twelve experiments were used without any knowledge transfer. 

The trained model can then be used in a variety of ways to determine the next process conditions that should be evaluated in a wet-lab experiment. In the simplest greedy scheme one could choose the process conditions for which the trained model predicts the highest expected \emph{titer}. More elaborate methods could utilise the uncertainty of the Gaussian Process regression model to balance exploitation and exploration. The optimal procedure for this has been a subject of research in computer science and statistics \cite{GPUCB1, GPUCB2, GPUCB_krause} but an algorithm tailored to bioprocess development requires future considerations. Thus concluding, we believe that the proposed algorithm will be able to significantly speed up bioprocess development, or will allow the gain in predictability to significantly improve the quality of such processes.

\bibliographystyle{naturemag}
\bibliography{references}

\appendix
\section{Appendix}
\subsection{Example of Predictions}

\begin{figure}[H]
\includegraphics[width=\textwidth]{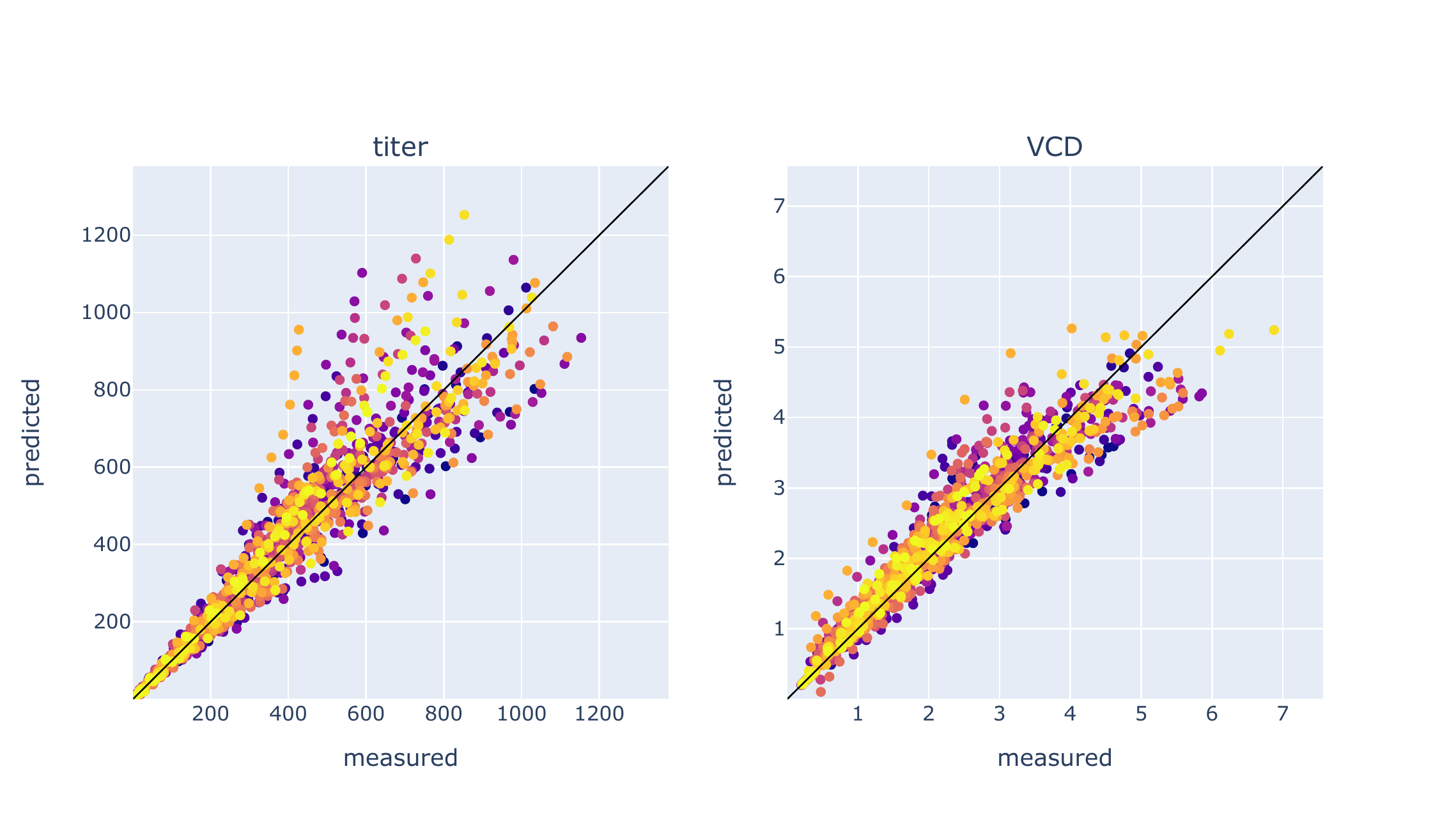}
\caption{Test set predictions after training with the example data set which is used for Figures 1a and 5 in the main text. Points with the same color come from the same experimental design at different time points.}
\label{fig:scatter_example}      
\end{figure}

\subsection{Distributions of Pairwise Distances for the Historic Products}
\begin{figure}[H]
  \includegraphics[width=\textwidth]{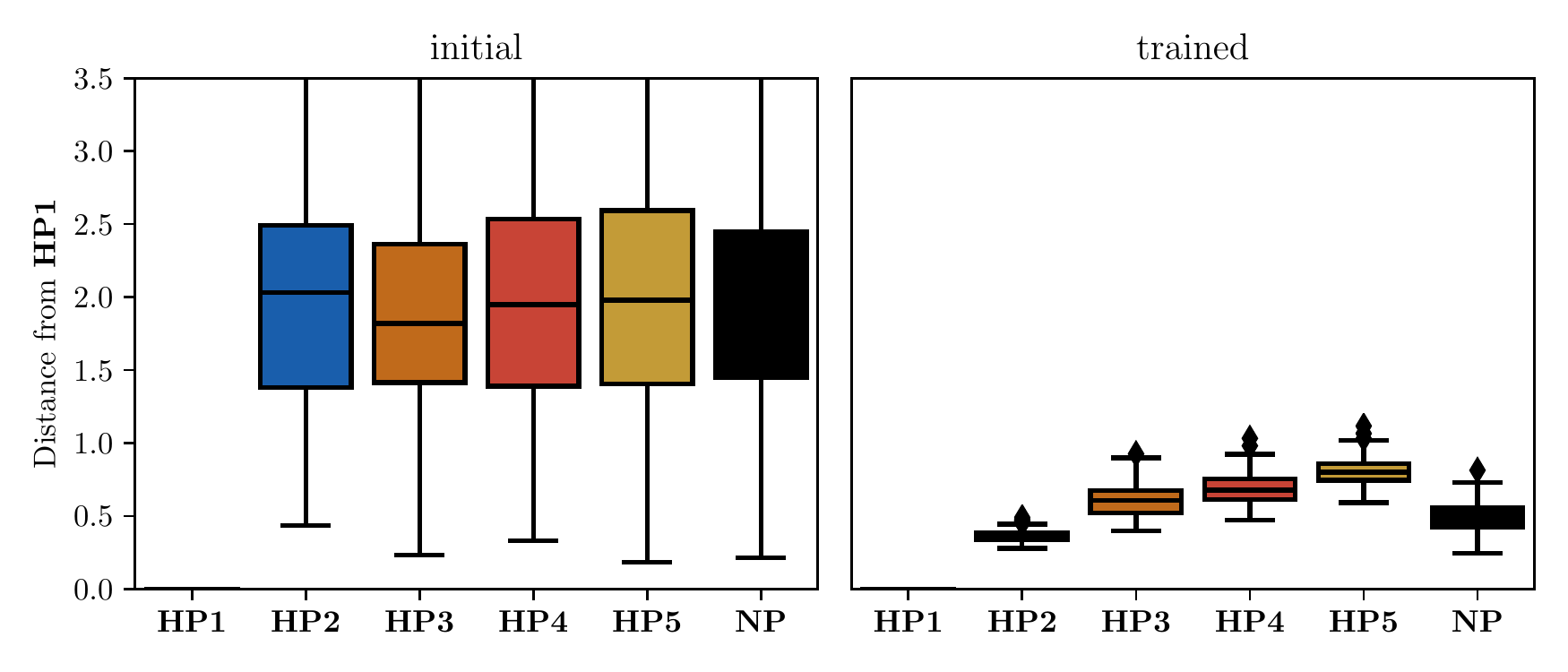}
\caption{Distribution of distances from the product \histProduct{1} across 150 repetitions. 
}
\label{fig:dist_HP1}       
\end{figure}

\begin{figure}[H]
  \includegraphics[width=\textwidth]{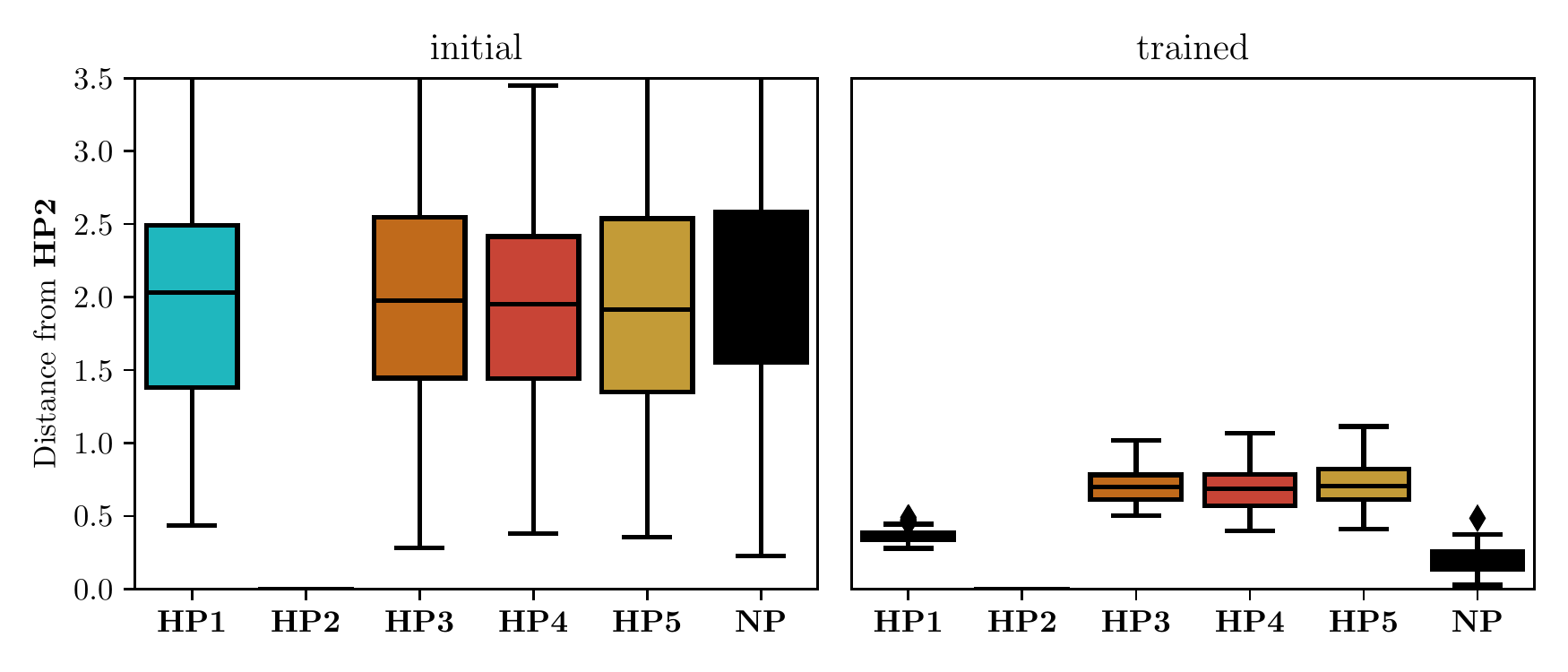}
\caption{Distribution of distances from the product \histProduct{2} across 150 repetitions. \histProduct{2} is between \novelProduct{} and \histProduct{1}.
}
\label{fig:dist_HP2}       
\end{figure}

\begin{figure}[H]
  \includegraphics[width=\textwidth]{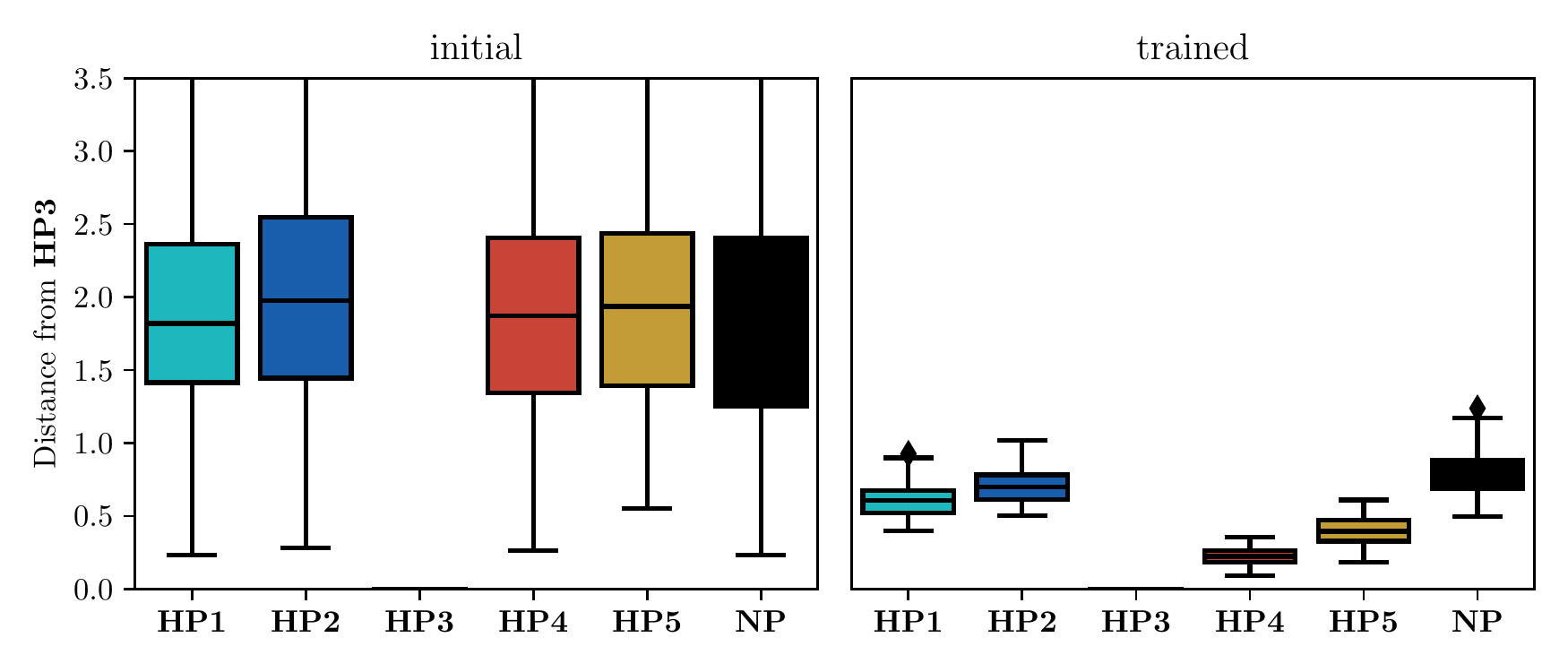}
\caption{Distribution of distances from the product \histProduct{3} across 150 repetitions.}
\label{fig:dist_HP3}       
\end{figure}

\begin{figure}[H]
  \includegraphics[width=\textwidth]{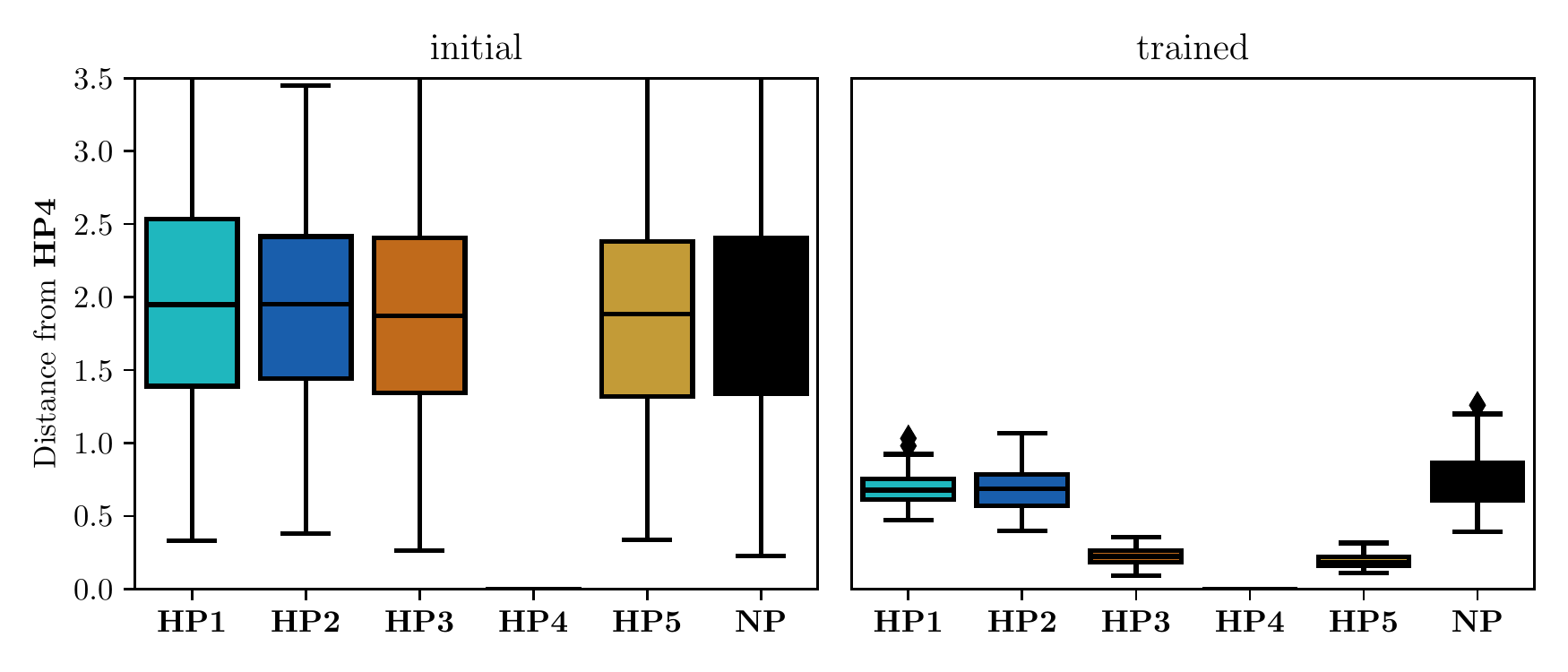}
\caption{Distribution of distances from the product \histProduct{4} across 150 repetitions. The product is between \histProduct{3} and \histProduct{5}.
}
\label{fig:dist_HP4}       
\end{figure}

\begin{figure}[H]
  \includegraphics[width=\textwidth]{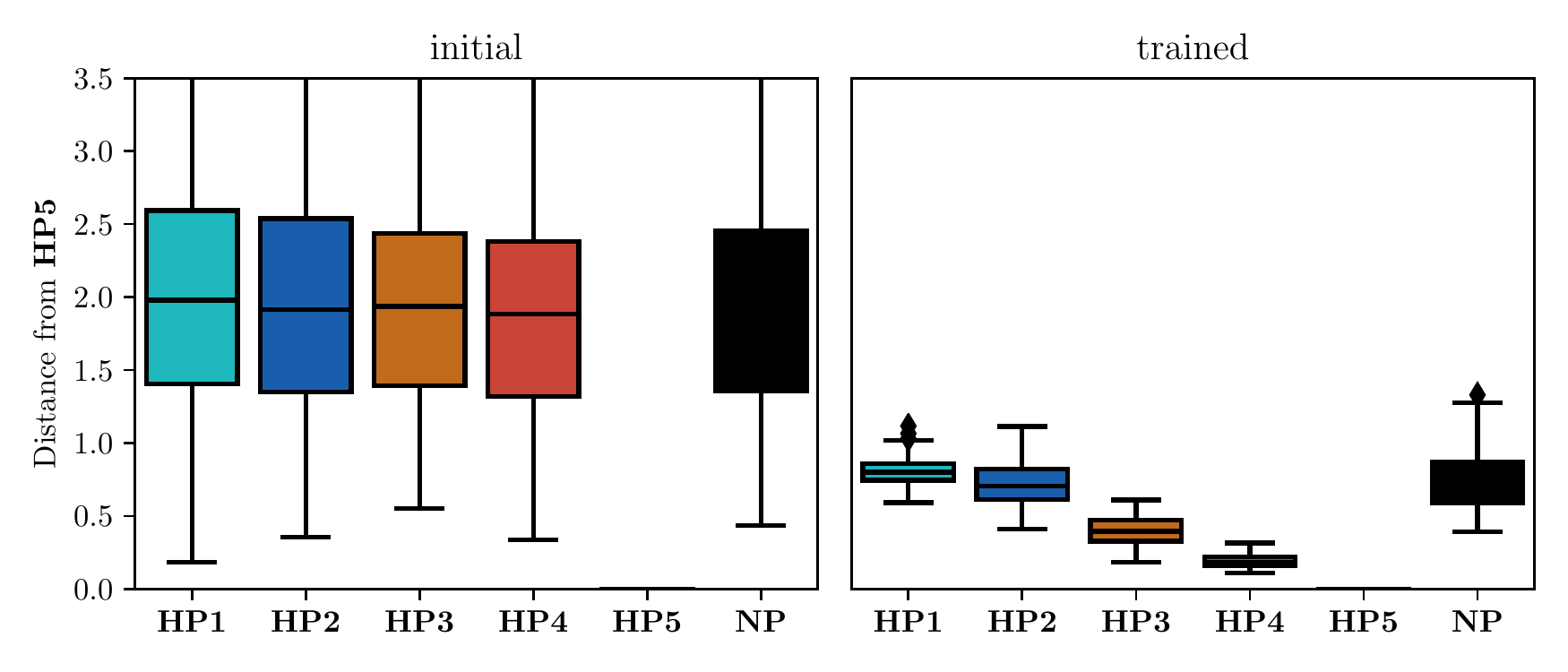}
\caption{Distribution of distances from the product \histProduct{5} across 150 repetitions.
}
\label{fig:dist_HP5}       
\end{figure}

\end{document}